 \documentclass[preprint]{aastex}

\slugcomment{}

\begin{document}


\title{A Broad 22 $\mu$m Emission Feature in the \\
    Carina Nebula \ion{H}{2} Region\altaffilmark{1}}


\author{Kin-Wing Chan and Takashi Onaka}
\affil{Department of Astronomy, School of Science, University of Tokyo,\\
    Bunkyo-ku, Tokyo 113-0033, Japan}
\email{kwc@astron.s.u-tokyo.ac.jp, onaka@astron.s.u-tokyo.ac.jp}


\altaffiltext{1}{ Based on observations with ISO, an ESA project with
instruments funded by ESA members states (especially the PI countries France, 
Germany, the Netherlands, and the United Kingdom) and with the participation 
of ISAS and NASA.}


\begin{abstract}
We report the detection of a broad 22 $\mu$m emission feature in the Carina
nebula \ion{H}{2} region by the Infrared Space Observatory (ISO) Short 
Wavelength Spectrometer. The feature shape is similar to that of the 22 $\mu$m
emission feature of newly synthesized dust observed in the Cassiopeia A 
supernova remnant. This finding suggests that both of the features are arising 
from the same carrier, and that supernovae are probably the dominant production 
source of this new interstellar grain. A similar broad emission dust feature is 
also found in the spectra of two starburst galaxies from the ISO archival data. 
This new dust grain could be an abundant component of interstellar grains and
can be used to trace the supernova rate or star formation rate in external
galaxies. The existence of the broad 22 $\mu$m emission feature complicates the
dust model for starburst galaxies and must be taken into account correctly in 
the derivation of dust color temperature. Mg protosilicate has been suggested 
as the carrier of the 22 $\mu$m emission dust feature observed in 
Cassiopeia A. The present results provide useful information in studies on
chemical composition and emission mechanism of the carrier.
\end{abstract}




\section{Introduction}
Supernovae have been suggested besides evolved stars as one of the major sources
 of interstellar dust (see Gehrz 1989, Jones and Tielens 1994, Dwek 1998 for 
review). Supporting evidence includes observations of dust condensation in the 
ejecta of SN 1987A (Moseley et al. 1989, Whitelock et al. 1989, Dwek et al. 
1992, Wooden et al. 1993), and those of the newly synthesized dust in the 
Cassiopeia A (Cas A) supernova remnant (Arendt, Dwek, \& Moseley 1999). The dust
 formation mechanism and the amount of dust that is formed in supernovae are
still poorly known. Observations of SN 1987A and Cas A showed that the mass of
the newly formed dust is much less than expected, and the discrepancy may be 
due to the fact that most of the dust is cold and cannot be detected in the 
far-infrared (Dwek 1998, Arendt et al. 1999). Finding an abundant dust 
component in the interstellar medium (ISM) which is formed only in supernovae 
will support the hypothesis that supernovae are a major source of interstellar 
dust. Furthermore, since the amount of this specific grain is proportional to 
the number of supernova, its total mass in the ISM can be used as a tracer of
the supernova rate or star formation rate in external galaxies. In this Letter 
we report the detection of a broad 22 $\mu$m emission dust feature in the
Carina nebula \ion{H}{2} region by the  ISO guaranteed time observations. We 
found that the shape of the present 22 $\mu$m emission dust feature is similar
to the 22 $\mu$m emission feature observed in Cas A. We also found a similar 
emission feature in two starburst galaxies from the ISO archival data.

\section{Observations}
The observations were made as part of the ISO guaranteed time program 
(TONAKA.WDISM1) using the Short Wavelength Spectrometer (SWS; de Graauw et al.
 1996). All the observations were made with the SWS AOT01 mode with scan speed
 of 1, which provided full grating spectra of 2.38 to 45.2 $\mu$m with a
resolution of $\lambda$/$\Delta$$\lambda$ = 300. The data have been processed
through the Off-Line Processing (OLP) 8.4, and reduced with the SWS Interactive
Analysis (IA) package developed by the SWS Instrument Dedicated Team. 
We observed the Car I \ion{H}{2} region in the Carina nebula and regions away
from it to the nearby molecular clouds (see de Grauuw et al. 1981 for
discussions of molecular clouds in the Carina nebula). The Car I \ion{H}{2}
 region is excited by the Trumpler 14, an open cluster containing numerous
 O-type stars. Totally four positions were observed. Pos 1 is at the Car I
\ion{H}{2} region with ${l}$ = 287$^{\circ}$.399 and ${b}$ = --0$^{\circ}$.633.
 Pos 2 (${l}$ = 287$^{\circ}$.349 and ${b}$ = --0$^{\circ}$.633), Pos 3 (${l}$
 = 287$^{\circ}$.299 and ${b}$ = --0$^{\circ}$.633), and Pos 4 (${l}$ =
 287$^{\circ}$.249 and ${b}$ = --0$^{\circ}$.633) are at a distance of 2.4,
 4.7, and 7.1 pc away from Pos 1, respectively.  Throughout our Letter, we
adopt a Sun-to-Carina nebula distance of 2.7 kpc (Grabelsky et al. 1988). 
Since the SWS aperture size varies across the wavelength ranges, we adjusted 
the difference in fluxes at the SWS band boundaries by scaling the spectra 
to the shortest band. The above adjustment does not affect the results
presented here.




\section{Results}
Figure 1a shows the observed SWS spectrum of the Carina nebula at Pos 1. A 
broad feature from $\sim$ 18 to 28 $\mu$m is clearly seen in the spectrum. 
The adjustment of the observed fluxes due to the different aperture sizes of
 SWS has no effect on this feature, since the SWS has the same aperture size
 from 12 -- 27.5 $\mu$m. It is difficult, however, to derive the spectral 
shape of this feature correctly since the underlying continuum emission is
very strong. We derived the feature shape by assuming the feature starts at 
18 $\mu$m and ends at 28 $\mu$m. Then the assumed underlying continuum 
emission, as shown in Figure 1a by the dashed line, is subtracted from the
 observed spectrum. The continuum emission comprises grains of graphite with
 temperature of 157 K and silicate with temperature of 40 K. Dust optical
 constants are adopted from Draine (1985).
The resultant feature shape is shown in Figure 1b, in which a peak around 
22 $\mu$m is clearly seen. This new 22 $\mu$m feature is distinctly 
different than the 21 $\mu$m feature that was discovered 
by Kwok, Volk, \& Hrivnak (1989). The 21 $\mu$m feature, 
which was only observed in carbon-rich post asymptotic
giant branch stars, has a much narrow feature width
of $\sim$ 4 $\mu$m (Volk, Kwok, \& Hrivnak 1999) compared to
that of the present 22 $\mu$m feature (with width of $\sim$ 10 $\mu$m).
This suggests that the 21 and 22 $\mu$m emission features 
are arising from different kinds of dust grain. 
Figure 2a shows another Carina nebula spectrum at Pos 2. The broad feature
 from 18 to 28 $\mu$m is also seen in this spectrum. The continuum emission
 comprising graphite with temperature of 135 K and silicate with temperature 
of 42 K is assumed (the dashed line in Fig. 2a) and subtracted from the
 observed spectrum. The excess emission is shown in Figure 2b. The unidentified
 infrared (UIR) emission features become stronger at Pos 2, a position farther
 away from the Car I \ion{H}{2} region compared to Pos 1. The slight difference
 in feature shape between figures 1b and 2b is probably due to the dust
 temperature effect. The 22 $\mu$m emission feature is also seen in Pos 3 and
 Pos 4 (not shown) but in weaker intensity. 

The same broad feature has been reported in the SWS spectra of M17--SW
 \ion{H}{2} region by Jones et al. (1999). They found in their spectra that
the intensity of this emission feature decreases with distance from the
 exciting stars, the same phenomenon we see in the present four observed
 spectra. The decrease of feature intensity may be due to: (1) dilution by the
 cool dust emission from the nearby molecular clouds; (2) emission of the
 feature requiring very high UV radiation intensity to be excited; and/or (3)
 decrease in the abundance of this specific grain with distance from the
 exciting stars. Identification of the carrier of the feature will help us to
 understand the observed decrease in the feature intensity.

Very recently, a broad emission dust feature with peak at 22 $\mu$m was 
reported in Cas A (Arendt et al. 1999). We compare this 22 $\mu$m feature 
and the present feature to see whether there is a similarity in feature shape.
 The comparison is shown in Figure 3. The Cas A spectrum was observed in the
 optical knot called N3 (see Arendt et al. 1999 for details), and is obtained
 from the ISO archival data. In order to obtain a better fit at wavelengths
 longer than 28 $\mu$m, we choose a new continuum emission, as shown in Figure
 1a by the dotted line, to give the 22 $\mu$m feature more long wavelength
 emission. The new continuum emission comprises graphite with temperature of
 160 K and silicate with temperature of 45 K. In Figure 3 we can see that the
 feature present in the Carina nebula shows a good agreement with that
 observed in Cas A. The origin of the excess emission around 13 $\mu$m is
 unknown. It should be noted, however, that the emission in Cas A at 
wavelengths between 20 and 50 $\mu$m may arise mostly from a warm ($\sim$ 
90 K) silicate component that originates from the diffuse shell (see Tuffs 
et al. 1999 for discussions of the spectral energy distribution of Cas A). 
If this warm silicate component is subtracted from the Cas A N3 spectrum, 
the resultant 22 $\mu$m feature (without emission at wavelengths longer than
 30 $\mu$m) will give a good fit to our observed 22 $\mu$m feature shown 
in Figure 1b.


\section{Discussion}
Evolved stars and supernovae have been suggested as the major production 
sources of interstellar dust. Past observations of evolved stars have found
 a number of dust features in the near to far-infrared ranges (see Waters et 
al. 1999 for a recent review). However, the broad 22 $\mu$m emission feature
 that we found in Carina nebula \ion{H}{2} region has never been reported in
 evolved stars. On the other hand, the present broad 22 $\mu$m emission 
feature is quite similar to the emission feature of newly synthesized dust
 observed in Cas A, suggesting that both of these features arise from the 
same dust grain, and that supernovae are probably the major production source
 of this new interstellar grain. The non-detection of the 22 $\mu$m feature
in SN 1987A (Moseley et al. 1989) does not make the latter suggestion less
 convincing, since the infrared emission in SN 1987A probably arises from
 optically thick clumps. Lucy et al. (1991) and
Wooden et al. (1993) suggest that the infrared emission in SN 1987A is
dominated by the dust in the optically thick clumps, and the low
density small grains in the interclump medium contribute
to the visual extinction. With this model, the infrared
emission in SN 1987A is a graybody emission, but the visual 
extinction is not.  

We would expect to find the 22 $\mu$m dust feature in astronomical sources
 with high supernova rate if supernovae are the major production source of
 this new interstellar grain.  Starburst galaxies are an ideal place to 
search for. From the ISO archival data we found that two starburst galaxies,
M82 and NGC7582, show a similar 22 $\mu$m emission feature. Figure 4 shows 
the SWS spectrum of the nuclear region of NGC7582, a narrow-line  X-ray 
galaxy with strong starburst in the central kpc (Radovich et al. 1999, 
and references therein). The 20 to 30 $\mu$m emission is mostly or 
completely arising from the broad 22 $\mu$m emission feature. The spectrum
of NGC7582 was taken by the SWS AOT01 with the speed of 2. We processed the
data through the OLP 8.4 and reduced with the SWS IA package in a way similar
to the Carina nebula spectra. The feature intensity in M82 (not shown) is
 much weaker, about 10$\%$ of the 18 -- 28 $\mu$m emission if the continuum
is assumed to pass through the 18 and 28 $\mu$m data points. Two other 
starburst galaxies, NGC253 and Circinus may also have a 22 $\mu$m feature,
but they are further weak in intensity and more observations are needed to
 confirm it.

The findings of the 22 $\mu$m dust feature in \ion{H}{2} regions and 
starburst galaxies suggest that this new grain could be an abundant component
 of interstellar dust. If the amount of this interstellar grain in the ISM
 is supposed to be proportional to the number of supernovae, its total mass 
in the ISM can be used as a tracer of the supernova rate or star formation 
rate in external galaxies. Studies of a large sample of starburst galaxies 
are required to confirm the above relationship. Only a limited number 
of galaxies have been observed by the SWS full grating scan mode, and a
 statistically useful sample of starburst galaxies for this study is not
 available at present. Future space missions like the Space InfraRed
Telescope Facility (SIRFT) and Infrared Imaging Surveyor (IRIS), and the
 Stratospheric Observatory for Infrared Astronomy (SOFIA) are expected
 to provide the necessary data base.

The existence of this broad 22 $\mu$m emission feature complicates the
dust model used in the study of the spectral energy distribution of 
starburst galaxies. Dust grains like graphite, amorphous carbon, silicates,
 and polycyclic aromatic hydrocarbons may not be representative of all the 
dust properties in starburst galaxies. Particularly, this broad 22 $\mu$m
 emission feature could have significant effects in the derivation of the 
dust color temperature based on the 20 -- 30 $\mu$m photometric flux (e.g., 
the Infrared Astronomical Satellite 25 $\mu$m data) as well as the number 
counts of deep surveys in the infrared spectral range to be carried out by
 SIRTF and IRIS observations, and must be taken into account appropriately.   

Arendt et al. (1999) suggested that the carrier of the 22 $\mu$m feature
 observed in Cas A is Mg protosilicate based on the good agreement between
the observed feature shape and the laboratory spectrum of the Mg 
protosilicate taken by Dorschner et al. (1980). They found that FeO can
also give a good fit to their observed 22 $\mu$m feature, but the required 
dust temperature higher than expected and the deficient of emission at
 wavelegths longer than 30 $\mu$m led them to rule it out as a promising
 candidate. If the identification of Mg protosilicate is true, it is the
 second silicate grain besides the astronomical silicates found in the ISM. 
More observations are needed to confirm (or test) the suggested identification.
 Observing the 22 $\mu$m feature in a variety of astronomical environments will
 provide useful information in studies on chemical composition and emission
 mechanism of the carrier.

The major results of this Letter are: (1) a broad 22 $\mu$m emission dust
 feature is detected in \ion{H}{2} regions and starburst galaxies; (2) the 
22 $\mu$m emission feature is similar in shape with the emission feature 
of newly synthesized dust observed in the ejecta of Cas A, and both of 
these features arise from the same carrier; and (3) supernovae are probably 
the major production source of this new interstellar dust.

\acknowledgments
We would like to thank the SWS IDT for providing the SWS IA software, and 
ISO project members for their efforts and help. We thank Robert Gehrz for 
useful comments. We also thank Issei Yamamura for useful discussions on the
 data reduction, and K. Kawara, Y. Satoh, H. Okuda, and the Japanese ISO team
 for their continuous help and encouragement.  K. W. C. is supported by the
 JSPS Postdoctoral Fellowship for Foreign Researchers. This work was 
supported in part by Grant-in-Aids for Scientific Research from JSPS.





\clearpage



\figcaption{(a) The observed SWS spectrum of the Carina nebula at Pos 1. The
 dashed (dotted) line represents the continuum emission with graphite and
 silicate at 157 (160) and 40 (45) K, respectively. The spectrum was scaled
 to the shortest band in order to adjust the difference in the aperture size.
 The flux unit is Jansky (Jy) per beam with beam size of \(14''\) $\times$
 \(20''\) in the shortest band. (b) The resultant 22 $\mu$m feature after
 subtraction of the dashed line continuum emission.  \label{fig1}}

\figcaption{(a) The observed SWS spectrum at Pos 2. The dashed line
 represents the continuum emission with graphite and silicate at 135 and
 42 K, respectively. (b) The resultant 22 $\mu$m feature after subtraction
 of the continuum.  \label{fig2}}

\figcaption{The Cas A N3 spectrum (the solid line) multiplied by a factor 
 5.2, together with the 22 $\mu$m feature (the + symbol) after subtraction of
 the dotted line continuum emission shown in Figure 1(a). For easy 
 comparison, the Cas A spectrum has been smoothed to a resolution of
 $\lambda$/$\Delta$$\lambda$ = 100. \label{fig3}} 

\figcaption{The SWS spectrum of NGC7582, which has been smoothed to a
 resolution of $\lambda$/$\Delta$$\lambda$ = 300.  \label{fig4}}


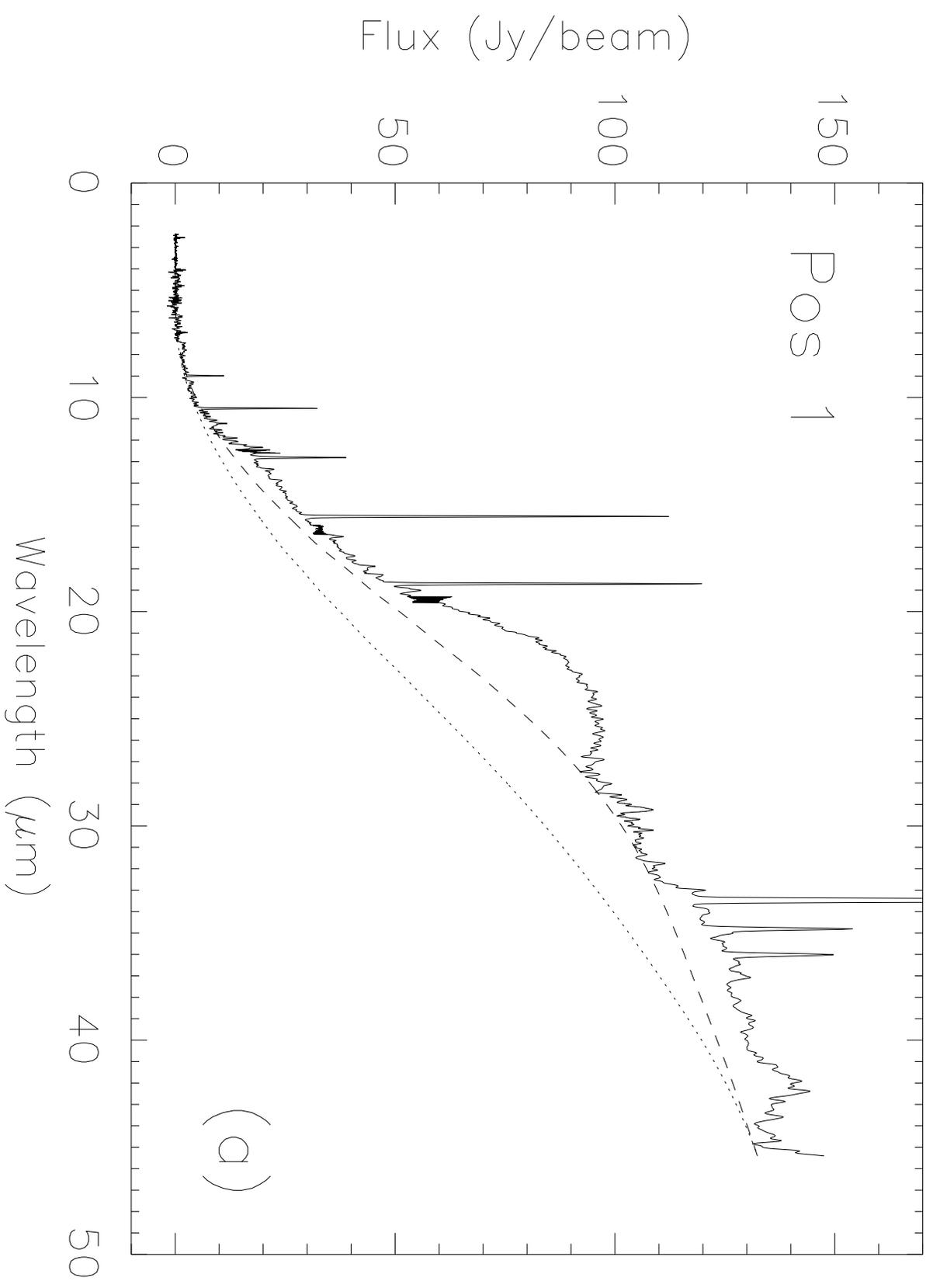
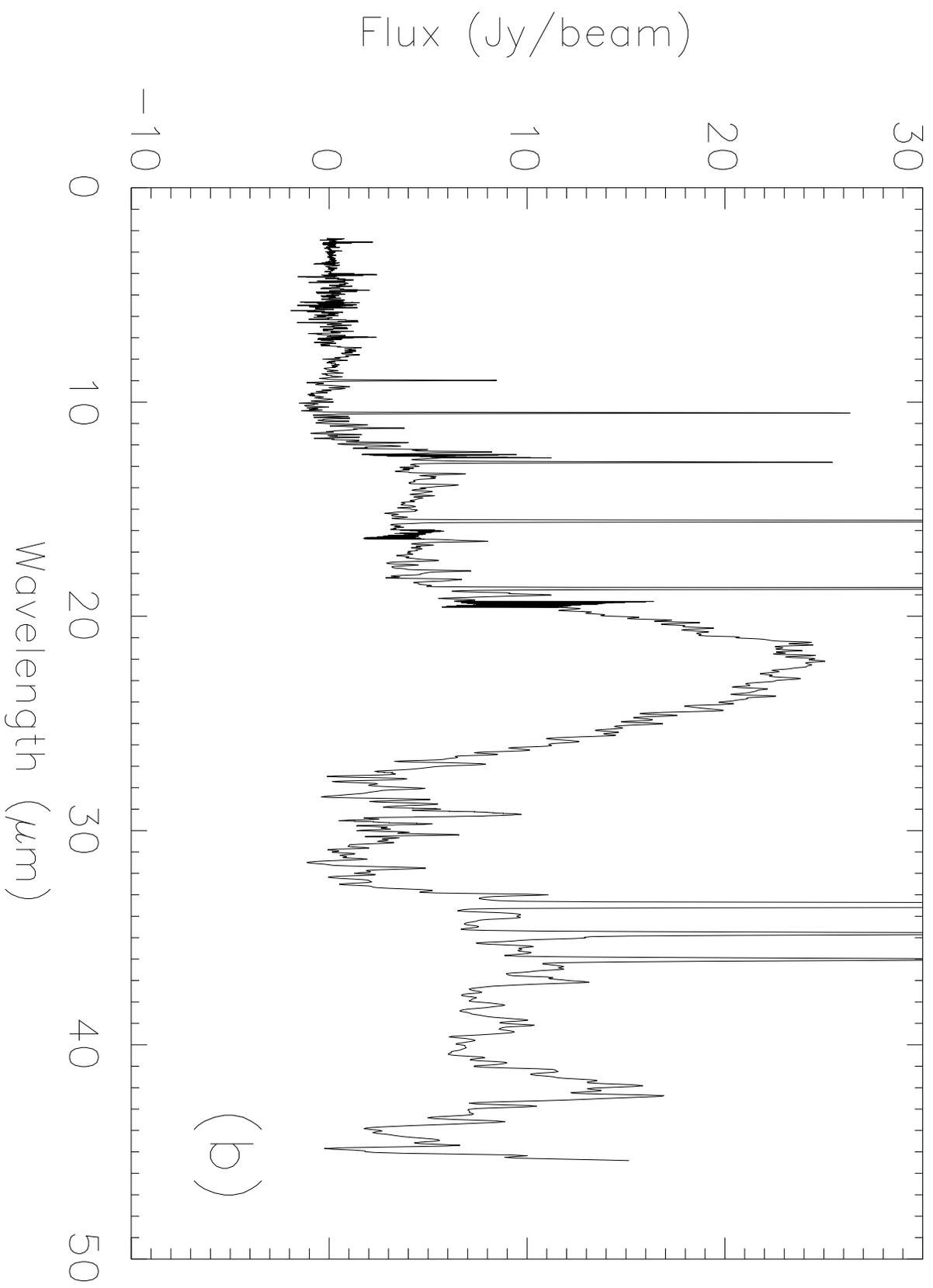
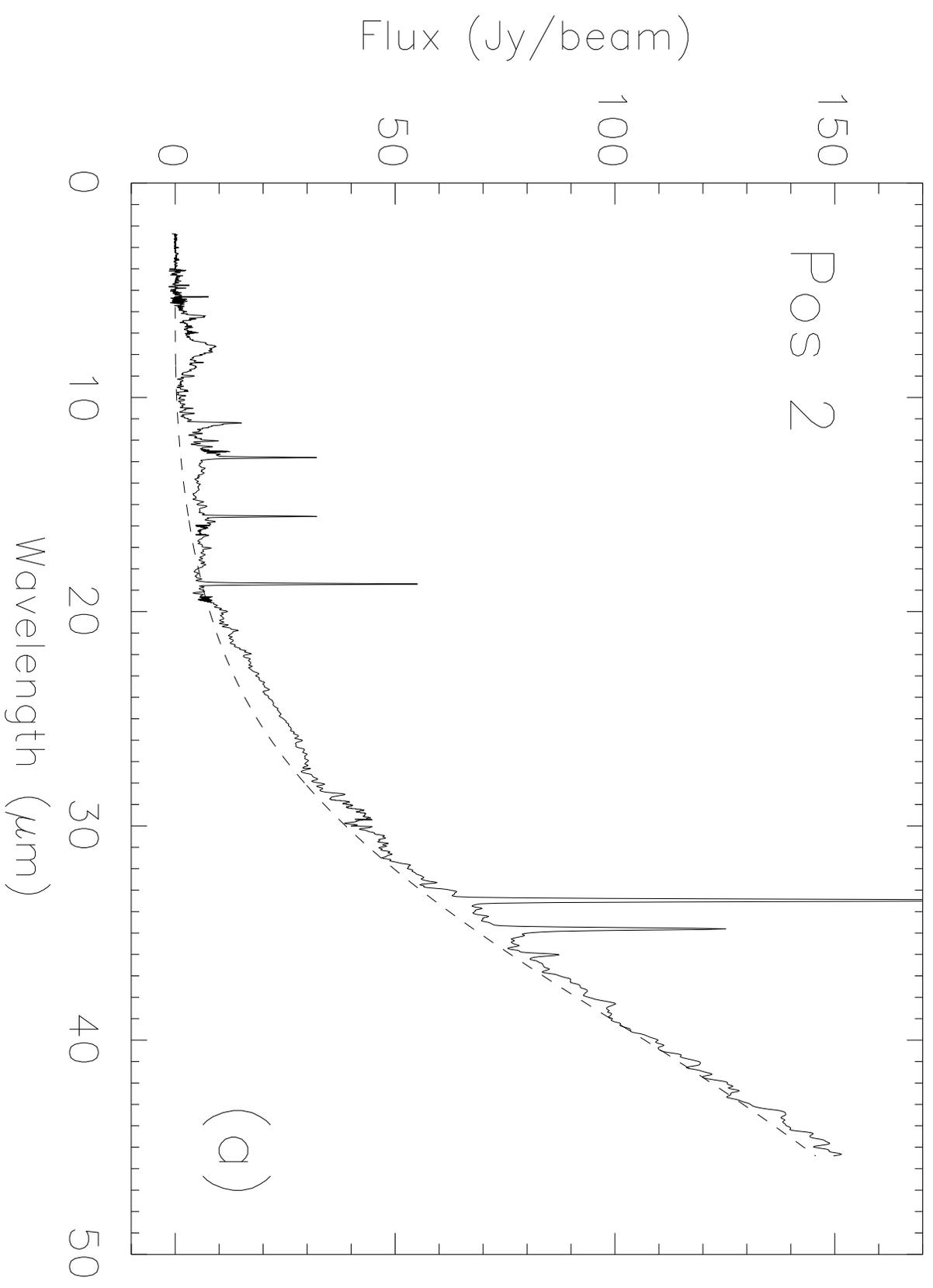
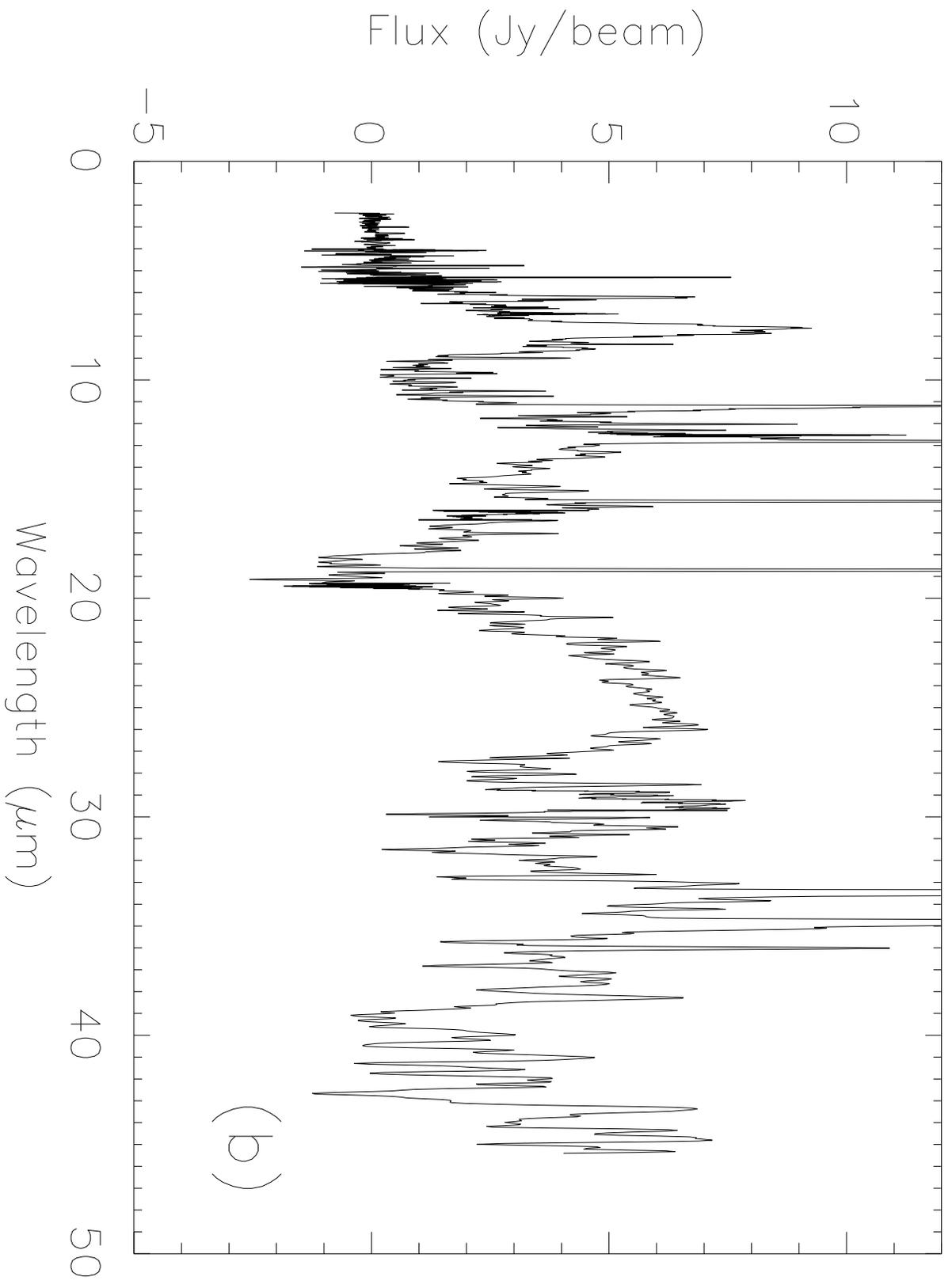
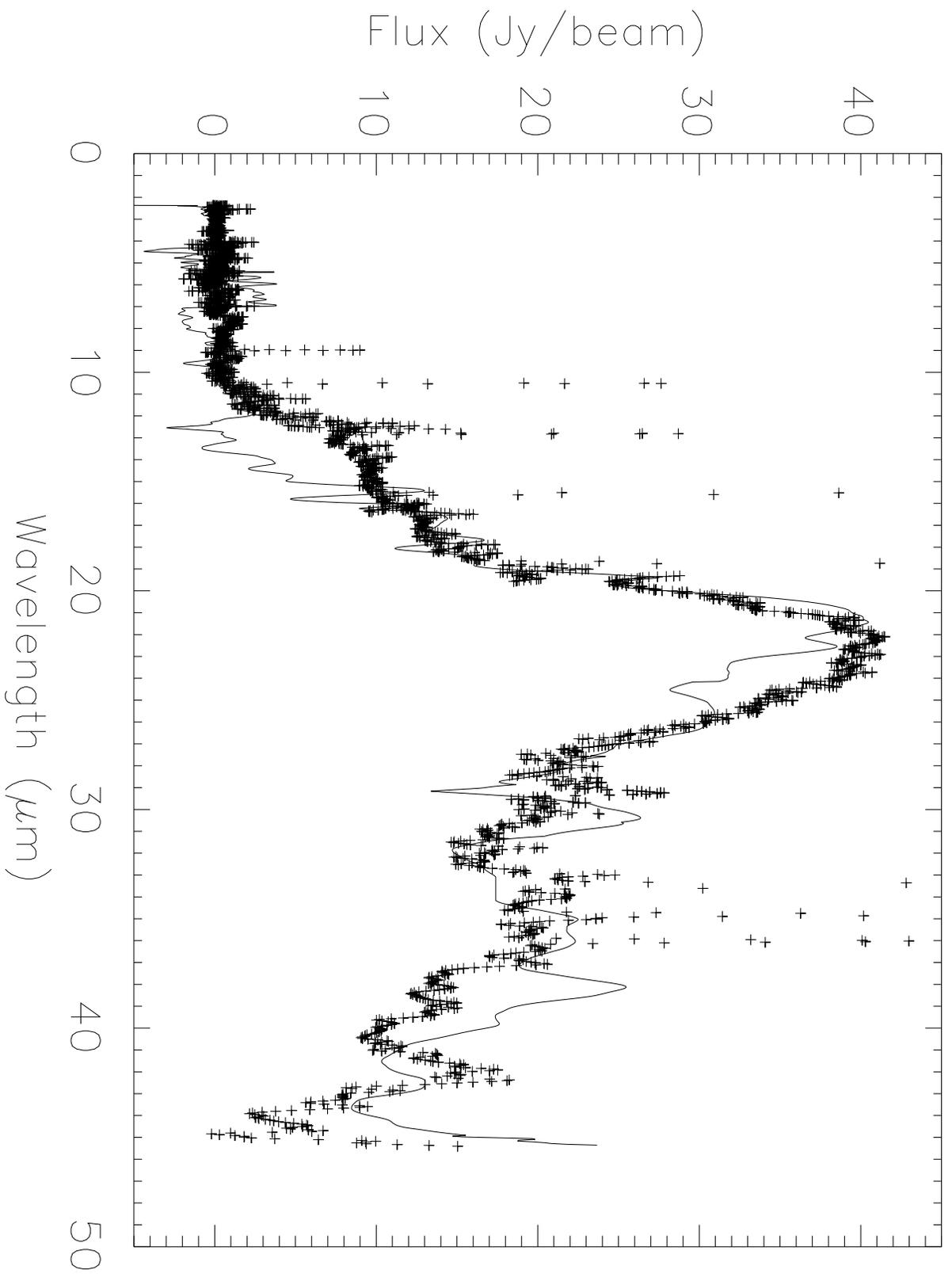
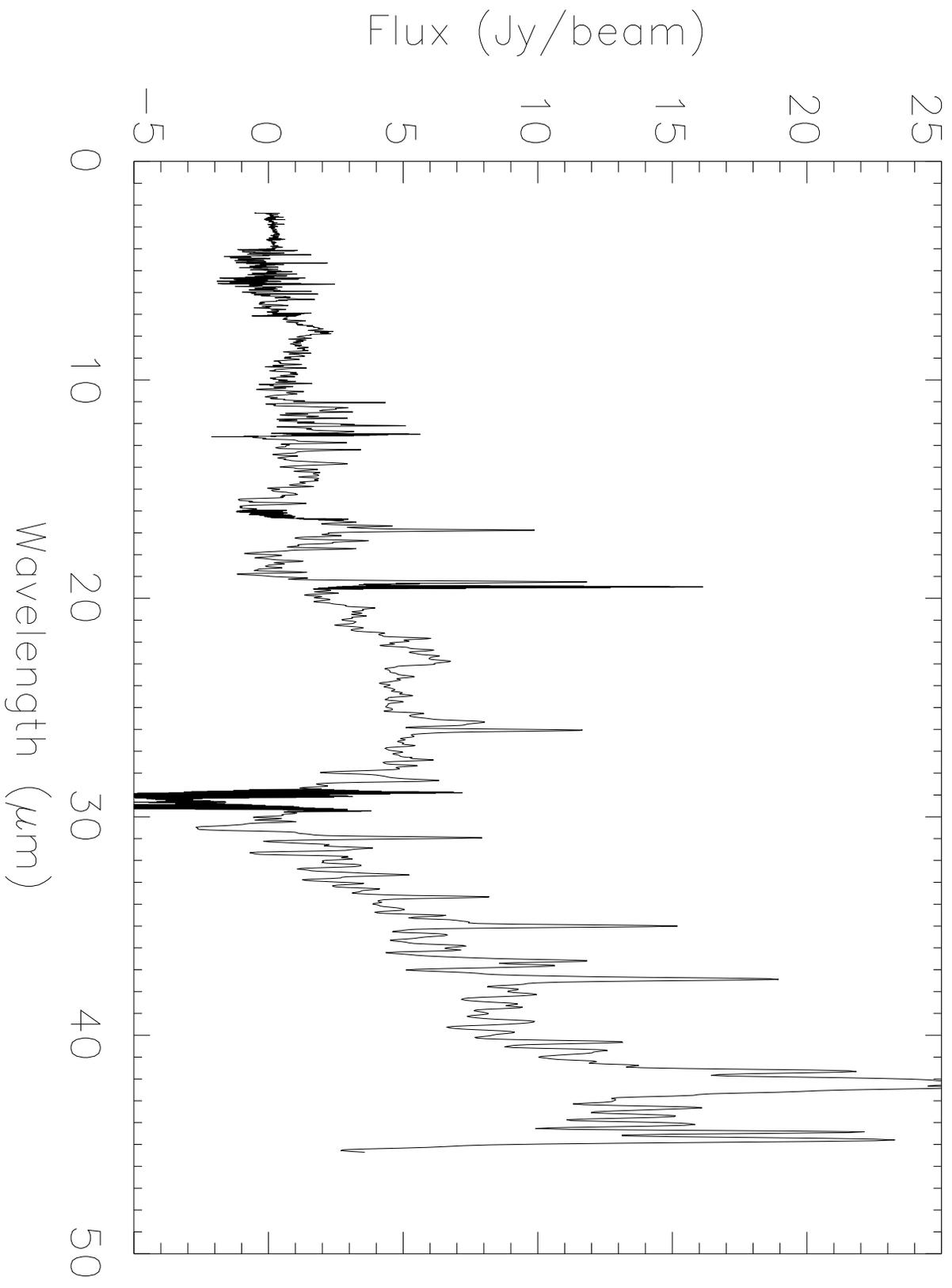
\end{document}